\documentclass[twoside]{article}
\usepackage[dvipsnames]{xcolor}
\usepackage{graphicx}
\usepackage{amsthm}
\usepackage{amsmath}
\usepackage{mathtools}
\usepackage{mathrsfs}
\usepackage{amssymb}
\usepackage{lipsum}
\usepackage{etoolbox}
\usepackage[linesnumbered,lined,boxed,ruled]{algorithm2e}
\usepackage[section]{placeins}
\usepackage[breaklinks,colorlinks,citecolor=blue,linkcolor=blue,urlcolor=blue]{hyperref}
\usepackage[all]{hypcap}
\usepackage{cancel}
\usepackage{csquotes}
\usepackage[toc,page]{appendix}
\usepackage{multirow}
\usepackage{datetime}
\usepackage{times}
\usepackage[format=plain, labelfont={bf,it}, textfont=it]{caption}
\usepackage{tcolorbox}
\usepackage{natbib, bibentry}
\usepackage{orcidlink}
\usepackage{fancyhdr}

\usepackage{scalerel}
\usepackage{stackengine,wasysym}

\newdateformat{mdydate}{\monthname[\THEMONTH] \THEDAY, \THEYEAR}
\newdateformat{monthyeardate}{\monthname[\THEMONTH] \THEYEAR}
\newdateformat{vmonthyeardate}{\THEYEAR.\THEMONTH.\THEDAY}

\newcommand{\be}{\begin{equation}}
\newcommand{\ee}{\end{equation}}
\newcommand{\bea}{\begin{eqnarray}}
\newcommand{\eea}{\end{eqnarray}}

\usepackage{geometry}
\graphicspath{{figures}}

\newcommand{\volume}{1}
\newcommand{\firstpage}{20}
\newcommand{\lastpage}{27}
\newcommand{\yyyy}{2024}
\newcommand{\mm}{October}
\newcommand{\dd}{21}
\newcommand{\authors}{Kruijssen et al.}
\newcommand{\fulltitle}{Merit-Based Sortition in Decentralized Systems}
\newcommand{\shorttitle}{Merit-Based Sortition in Decentralized Systems}

\newcommand{\doi}{10.70235/allora.0x\volume\ifnum\numexpr\firstpage<10 000\else\ifnum\numexpr\firstpage<100 00\else\ifnum\numexpr\firstpage<1000 0\fi\fi\fi\firstpage}
\fancypagestyle{firstpage}{%
    \fancyhead[L]{\footnotesize Allora Decentralized Intelligence \textbf{\volume}, \firstpage--\lastpage; \yyyy\ \mm\ \dd}
    \fancyhead[R]{\footnotesize doi:\href{https://doi.org/\doi}{\textcolor{blue}{\doi}}}
    \fancyfoot{}
    
}

\AtBeginDocument{\thispagestyle{firstpage}}

\begin{document}

\title{\fulltitle}
\author{\authors}
\date{\monthyeardate{\today}}

\vskip30mm
\begin{center}
\begin{minipage}{170mm}
\begin{center}
\vskip5mm
{\fontsize{15pt}{15pt}\textbf{\fulltitle}}
\vskip5mm
J.~M.~Diederik Kruijssen$^{\orcidlink{0000-0002-8804-0212}}$,
Renata Valieva$^{\orcidlink{0000-0002-7256-5321}}$,
Kenneth Peluso$^{\orcidlink{0009-0000-7299-6984}}$,
Nicholas Emmons$^{\orcidlink{0009-0009-6429-5921}}$ \&
Steven~N.~Longmore$^{\orcidlink{0000-0001-6353-0170},*}$
\vskip1mm
\textit{Allora Foundation}
\end{center}
\end{minipage}
\end{center}
\vspace{3mm}

\renewcommand{\thefootnote}{\fnsymbol{footnote}}
\footnotetext[1]{This author is faculty at Liverpool John Moores University. He co-authored this work in his separate capacity as an advisor to Allora Labs.}
\renewcommand{\thefootnote}{\arabic{footnote}}

\begin{abstract}
\noindent In decentralized systems, it is often necessary to select an `active' subset of participants from the total participant pool, with the goal of satisfying computational limitations or optimizing resource efficiency. This selection can sometimes be made at random, mirroring the sortition practice invented in classical antiquity aimed at achieving a high degree of statistical representativeness. However, the recent emergence of specialized decentralized networks that solve concrete coordination problems and are characterized by measurable success metrics often requires prioritizing performance optimization over representativeness. We introduce a simple algorithm for `merit-based sortition', in which the quality of each participant influences its probability of being drafted into the active set, while simultaneously retaining representativeness by allowing inactive participants an infinite number of chances to be drafted into the active set with non-zero probability. Using a suite of numerical experiments, we demonstrate that our algorithm boosts the quality metric describing the performance of the active set by $>2$ times the intrinsic stochasticity. This implies that merit-based sortition ensures a statistically significant performance boost to the drafted, `active' set, while retaining the property of classical, random sortition that it enables upward mobility from a much larger `inactive' set. This way, merit-based sortition fulfils a key requirement for decentralized systems in need of performance optimization.
\end{abstract}
\vspace{3mm}

\section{Introduction} \label{sec:intro}
The term `sortition' originally refers to the process of randomly selecting representatives in a democratic system, a practice dating back 2.5 millennia to ancient Athens, where the selection of public officials by lottery was seen as the best way of achieving fairness in society \citep[e.g.][]{headlam1891}. Since then, sortition has spread over the world as a way of obtaining representative selections of politicians, public officials, or advisors, going through ebbs and flows in terms of its popularity \citep[e.g.][]{flanigan21,jacquet22,sintomer23}. Advocates of sortition often highlight positive implications such as fairness, representativeness, and efficiency \citep[e.g.][]{engelstad89,sintomer23b}.

In permissionless, decentralized systems \citep[e.g.][]{nakamoto08,buterin14}, a form of sortition is often needed too. Across a large set of anonymous contributors, validators, or other participants, there exists a high degree of redundancy that does not require all participants to be involved in the decision process \citep[e.g.][]{wuest18}. Sometimes, there may exist concrete computational limitations why it is infeasible to involve all participants. In any of such cases, (pseudo)-random subsets may be able to fulfil the same task without a serious loss of security or performance. In principle, this can be done using classical, random sortition techniques (of which numerous trustless examples exist, see e.g.\ \citealt{gilad17,saa19,freitas23}) to achieve representativeness and efficiency.

However, recent developments in decentralized networks have brought about a rapid growth of systems that aim to achieve concrete goals, often with measurable performance or success. Examples are decentralized machine intelligence and inference systems \citep[e.g.][]{rao21,kruijssen24}, on-chain oracles \citep[e.g.][]{ellis17,breidenbach21}, or internet-of-things networks \citep[e.g.][]{banerjee23}. In such systems, which aim to achieve a quantifiable degree of success, the goal of sortition is to improve efficiency not only while maintaining representativeness, but also while optimizing performance. The latter goal is achieved not through random sortition, but by letting the performance of each participant influence their probability of being drafted. We refer to this concept as merit-based sortition.

In this paper, we introduce a simple algorithm for merit-based sortition that can be used to increase computational efficiency by limiting active participation without sacrificing (and generally improving) performance. This is possible, because the algorithm:
\begin{enumerate}
\item
optimizes the quality of the active set of participants by letting the probability of relegation out of the active set decrease with the participant's quality;
\item
retains fairness and representativeness by allowing inactive participants an infinite number of chances to be drafted into the active set, in such a way that the probability and frequency of promotion increase with the participant's quality.
\end{enumerate}
In \S\ref{sec:method} we outline the proposed algorithm for merit-based sortition. We investigate its quantitative performance and behavior in \S\ref{sec:sims} using a suite of numerical experiments. Finally, we summarize our results in \S\ref{sec:disc}.

\section{Algorithm for Performing Merit-Based Sortition} \label{sec:method}
Consider a system consisting of $N_{\rm tot}$ participants, which are required to repeatedly carry out a certain task over the course of many epochs (or time steps) $i\in\{1,\ldots,N_{\rm epochs}\}$, where $N_{\rm epochs}$ represents the total number of epochs. For some reason (e.g.\ limited resources or higher efficiency), only an `active' set $N_{\rm act}<N_{\rm tot}$ of the participants can be chosen to carry out this task. Contrary to classical sortition, where the active set is selected at random to optimize representativeness, we aim to select the participants that are best at carrying out the task. Over time, we wish to retain a degree of permeability to enable inactive participants to enter the active set and prove their worth. This permeability allows the algorithm to appropriately handle quality changes among the participants.

We assume that there exists a metric $Q_{ij}$, agreed on by vote, validation, or on-chain calculation within a decentralized system, that describes the quality of the work by participant $j\in \{1,\ldots,N_{\rm tot}\}$ in the system during epoch $i$. In the context of Allora, $Q_{ij}$ will typically reflect the score of an inferer, forecaster, or reputer, but it can correspond to any metric that expresses quality, performance, or ability. In order to select the active set based on their quality, it is natural to select the top $N_{\rm act}$ participants by ordering them by $Q_{\rm i-1,j}$, i.e.\ their quality metric during the previous epoch. This results in an active set ${\cal A}_i(Q_{\rm i-1,j})$ with size $N_{\rm act}$, and an inactive set ${\cal I}_i(Q_{\rm i-1,j})$ with size $N_{\rm inact} = N_{\rm tot}-N_{\rm act}$.

A simple definition of the active and inactive sets based on the instantaneous quality metrics at epoch $i-1$ runs into two key issues:
\begin{enumerate}
    \item Quality metrics are likely to be volatile, and $Q_{i-1,j}$ may be a poor predictor for the expectation value at the current epoch $E(Q_{ij})$.
    \item Making the reasonable assumption that the quality metric is assigned based on actual performance during the previous epoch, then it is unavailable for the inactive participants, even if the participant might have been active at an earlier epoch $i'$. The distribution of quality metrics at $i'$ may have been different from that at $i-1$, making $Q_{i'j}$ unsuitable for use in the ranking. This means that the active set is impermeable, because inactive participants cannot be ranked without assigning them an appropriate quality metric.
\end{enumerate}

We now show that both of these two problems can be addressed by applying an exponential moving average (EMA) to smoothen the quality metric and updating the EMA of the inactive participants with an appropriate dummy value that is derived from the active set ${\cal A}_i$. At epoch $i$, the smoothed quality metric of participant $j$ is defined as:
\be
\label{eq:q_ema}
Q_{ij} = \alpha \hat{Q}_{ij} + (1-\alpha)Q_{i-1,j} ,
\ee
where $\alpha\in(0,1]$ represents the EMA smoothing factor (higher values imply less smoothing and more recency bias) and $\hat{Q}_{ij}$ is the target quality metric used to update the EMA. For the active set of participants ${\cal A}_i$, we define
\be
\label{eq:q_target_active}
\hat{Q}_{ij} = T_{ij}~{\rm for}~j\in{\cal A}_i ,
\ee
where $T_{ij}$ is the instantaneous quality metric (e.g.\ score) of participant $j$ during epoch $i$. Naturally, $T_{ij}$ only exists for the active set of participants. This addresses the first of the above two issues. If the instantaneous quality metric is highly volatile and $T_{i-1,j}$ is a poor predictor of $E(T_{ij})$, then taking the EMA ensures that $Q_{i-1,j}$ is a better predictor of $E(Q_{ij})$. This makes $Q_{i-1,j}$ more suitable for defining ${\cal A}_i$ and ${\cal I}_i$ than $T_{i-1,j}$.

The second question is now how to define $\hat{Q}_{ij}$ for the inactive set of participants ${\cal I}_i$, because their inactivity means they do not possess an associated instantaneous quality metric $T_{ij}$. The definition of $\hat{Q}_{ij}$ determines the asymptote towards which the EMA tends. As such, it can be used to regulate the permeability of the separation between the active and inactive sets of participants. If $\hat{Q}_{ij}<\min_{\cal A}(T_{ij})$, where $\min_{\cal A}$ indicates taking the minimum over participants in the active set, then $\max_{\cal I}(Q_{ij})<\min_{\cal A}(Q_{ij})$, i.e.\ no member of the inactive set will ever achieve a quality metric high enough to become part of the active set. However, if $\hat{Q}_{ij}>\min_{\cal A}(T_{ij})$, then it is possible that $\max_{\cal I}(Q_{ij})>\min_{\cal A}(Q_{ij})$, i.e.\ inactive participants obtain a smoothed quality metric high enough to be included in the active set. The actual occurrence of such a promotion into the active set depends on:
\begin{enumerate}
    \item The difference between the smoothed quality metric of the inactive participant during the previous epoch $Q_{i-1,j}$ and the minimum instantaneous quality metric of the active participants $\min_{\cal A}(T_{ij})$. A comparatively higher value of $Q_{i-1,j}$ means that a smaller difference needs to be bridged to exceed $\min_{\cal A}(T_{ij})$, and thus increases the probability of a promotion. Likewise, a higher volatility of the instantaneous quality metrics implies a lower $\min_{\cal A}(T_{ij})$ and thereby increases the probability that a spot in the active set opens up.
    \item The smoothing factor $\alpha$. A higher value of $\alpha$ means that more weight is given to the target quality metric $\hat{Q}_{ij}$, causing the smoothed quality metric to change more rapidly, and thus increases the probability of a promotion.
    \item The percentile $P$ of $T_{ij}$ to which $\hat{Q}_{ij}$ corresponds. A higher percentile means that the target quality metric $\hat{Q}_{ij}$ exceeds $\min_{\cal A}(T_{ij})$ by a larger amount, and thus increases the probability of a promotion.
\end{enumerate}

We use the third of the above points to define $\hat{Q}_{ij}$ for the inactive set ${\cal I}_i$ as the percentile $0<P[\%]\leq100$ of the instantaneous quality metrics of the active set ${\cal A}_i$:
\be
\label{eq:q_target_inactive}
\hat{Q}_{ij} = f_{P}(T_{ij'})~{\rm for}~j\in{\cal I}_i~\land~j'\in{\cal A}_i ,
\ee
where $f_P$ is a function that calculates the percentile $P$ of its argument. The percentile $P$ defines what fraction of the active set gets relegated and renewed, and thus also controls the rate of cycling between the active and inactive sets. At any epoch $i$, the bottom $P$-th percentile of participants in ${\cal A}_i$ (ordered by $Q_{ij}$) are at risk of getting demoted, and will be replaced if any participants in ${\cal I}_i$ have sufficiently high $Q_{i-1,j}$, $\alpha$, $P$, and thus $\max_{\cal I}(Q_{ij})>\min_{\cal A}(Q_{ij})$. If $f_{P}(T_{ij'})$ is not instantly available, it is also possible to instead use the preceding percentile, i.e.\ $f_{P}(T_{i-1,j'})$ for $j\in{\cal I}_i~\land~j'\in{\cal A}_{i-1}$.

Finally, we need to handle the situation where a participant is part of the active set ${\cal A}_i$, but somehow is unable to contribute to epoch $i$ and thus does not produce an associated instantaneous quality metric $T_{ij}$. In this case, it is undesirable to treat the participant as if it were part of the inactive set ${\cal I}_i$ and apply \autoref{eq:q_target_inactive}, because this would make it remain in the active set due to its high EMA-smoothed quality metric. The solution is to set $\hat{Q}_{ij}$ as the minimum instantaneous quality metric (i.e.\ the zeroth percentile) with an additional penalty for neglecting to contribute. The penalty is well-expressed as a multiple of the standard deviation of the instantaneous quality metrics $T_{ij}$, because it controls the time needed to achieve promotion back into the active set:
\be
\label{eq:q_target_active_awol}
\hat{Q}_{ij} = {\rm min}_{\cal A}(T_{ij'})-\lambda_{\rm pen}\sigma_{\cal A}(T_{ij'})~{\rm for}~\{j,j'\}\in{\cal A}_i~\land~\nexists~T_{ij} ,
\ee
where $\sigma_{\cal A}$ indicates taking the standard deviation of the instantaneous quality metrics that were contributed by the active set at epoch $i$, and we adopt $\lambda_{\rm pen}=2$ for a $2\sigma$ penalty.

The above ruleset allows selecting the top $N_{\rm act}$ participants by ordering them by $Q_{\rm i-1,j}$, i.e.\ their quality metric during the previous epoch, without running into the issues outlined above. If the separation between the active and inactive sets intersects a subset of participants with equal $Q_{i-1,j}$, then a random subset of these equal-valued participants is drawn so that the active set contains exactly $N_{\rm act}$ participants. Similarly, when the number of participants that have an existing $Q_{i-1,j}$ from a previous epoch is smaller than the desired $N_{\rm act}$, further participants are drawn at random from any remaining (new) participants and added to the active set up to a total of $N_{\rm act}$ participants.

\section{Numerical Simulations} \label{sec:sims}
We perform a suite of numerical experiments to demonstrate the performance of the merit-based sortition algorithm described in \S\ref{sec:method}. Each participant is assigned a median quality metric, obtained by drawing from a normal distribution $\tilde{Q}_j\leftarrow{\cal N}(\mu_Q,\sigma_Q)$ with default mean $\mu_Q=0.2$ and standard deviation $\sigma_Q=0.1$. This median quality metric reflects the typical ability or performance of the participant. Each epoch, an instantaneous quality metric is generated for each participant (active or not) by drawing from a normal distribution as $T_{ij}\leftarrow{\cal N}(\tilde{Q}_j, V)$, where $V=0.2$ is a parameter representing the quality volatility. Each simulation starts with a set of $N_{\rm init}$ participants, after which new participants are generated each epoch by drawing from a Poisson distribution ${\cal P}$ as $N_{\rm growth}\leftarrow{\cal P}(p_{\rm growth})$, with default value $p_{\rm growth}=10^{-10}$ (almost no growth). Each participant is initialized with a total participation lifetime that is drawn from a geometric distribution ${\cal G}$ (representing the number of Bernouilli trials needed to achieve one success) as $\Delta t\leftarrow{\cal G}(p_{\rm attr})$ with default value $p_{\rm attr}=10^{-10}$ (almost no attrition). The default smoothing factor is $\alpha=0.1$ and each experiment is run for a total of $N_{\rm epochs}=1000$ epochs. The percentile $P$ and the numbers of initial and active participants vary between the experiments, and will be indicated in each of the figures below.

\subsection{Simple example} \label{sec:simple}
\begin{figure*}[tb]
\centering
\includegraphics[width=0.35\textwidth]{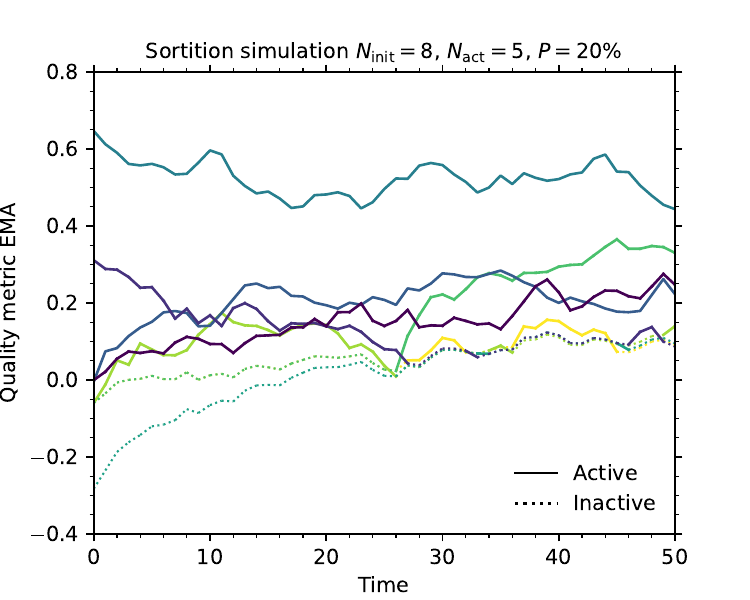}%
\hspace{-0.03\textwidth}
\includegraphics[width=0.35\textwidth]{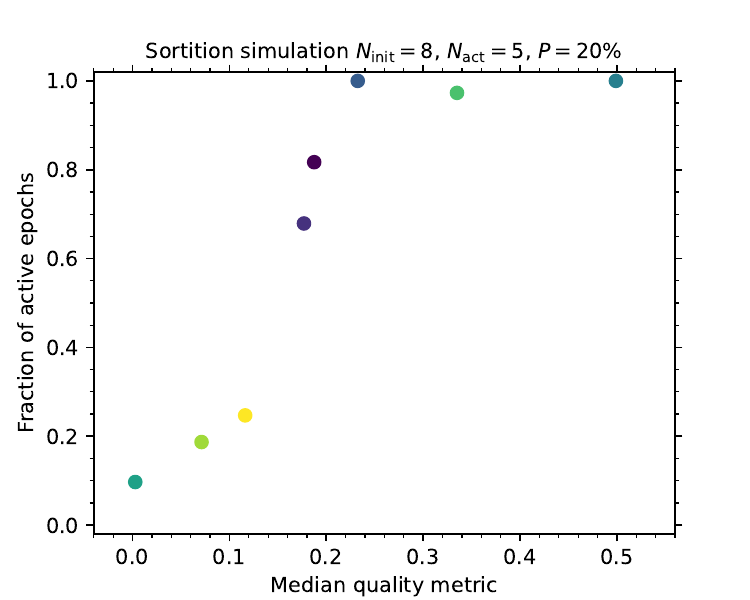}%
\hspace{-0.03\textwidth}
\includegraphics[width=0.35\textwidth]{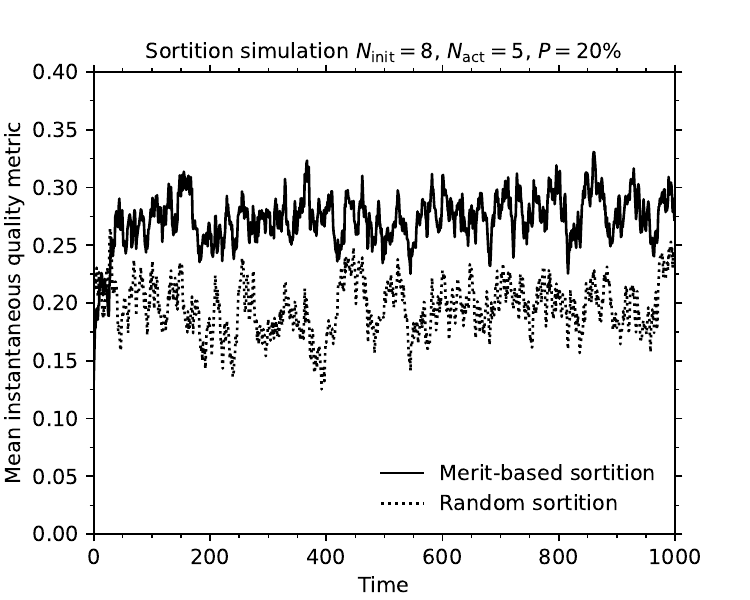}%
\caption{Merit-based sortition simulation for an initial set of eight participants, illustrating the construction and performance of the active set ${\cal A}_i$ as a function of time. Left: Each color represents a different participant, and the line style reflects the (in)activity as indicated by the legend. Line segments connecting two epochs where a participant changes from inactive to active show as solid, whereas those connecting a transition from active to inactive show as dotted. Middle: Correlation between the median quality metric of each participant (reflecting their ability or performance) and the fraction of the 1000-epoch duration of the simulation for which this participant is active. Colors match those in the left-hand panel. Right: EMA of the mean instantaneous quality metric across the active set ${\cal A}_i$ for merit-based sortition (solid line) and random sortition (dotted line) over the full duration of the simulation. The panel titles list the number of initial participants, the number of active participants, and the percentile.\label{fig:default}}
\end{figure*}
\autoref{fig:default} illustrates the typical behavior of merit-based sortition for the aforementioned default parameters, an initial set of $N_{\rm init}=8$ participants, a pool of $N_{\rm act}=5$ active participants, and a percentile value of $P=20\%$. The left-hand panel covers the first 50 epochs of the experiment and showcases a variety of relevant properties exhibited by the sortition scheme. During the very first epoch, a random draw is made to select five active participants out of the total eight. The three inactive participants are assigned a quality metric corresponding to the percentile $P$, which is about $-0.05$ at $i=0$. From this point emerges one solid line: this participant is promoted immediately, again through random selection as the three inactive participants have an indistinguishable quality metric EMA. Additionally, one dotted line emerges from $\{0, -0.05\}$, which represents the two remaining participants who are initially inactive -- these continue to overlap, because their quality metric EMA is built from the same percentile of the active participants. The bottom-most dotted line represents a participant who is active at $i=0$, but underperforms greatly and is relegated immediately. It subsequently slowly recovers, but remains queued behind the other two inactive participants.

After $i=26$, a participant who promoted after $i=0$ is relegated, allowing the first of the two remaining originally-inactive participants (selected at random) to get promoted. After $i=27$, the final originally-inactive participant gets promoted -- this time on merit, because there are no more identical participants. The participant relegated after $i=27$ only slightly underperformed, and therefore is the first to (briefly) be promoted again at $i=33$. As a result, it takes till after $i=33$ for the participant who was relegated at $i=0$ to be finally promoted back into the active set ${\cal A}_{34}$. This participant fails to prove its worth, is relegated again after one attempt, and gets another chance in the active set only after $i=45$. Three out of eight participants are active throughout the 50 epochs shown here. Unsurprisingly, these are among the ones with the highest median quality metric $\tilde{Q}_j$, even though the second-best participant was inactive at $i=0$ and thus had to wait for promotion before getting a first chance. After 1000 epochs, this participant has never relinquished its spot in ${\cal A}_i$ and has been active for 97.3\% of all epochs.

Indeed, the center panel in \autoref{fig:default} shows that there exists a clear correlation between each participant's median quality metric and the fraction of epochs that they are active. Likewise, the right-hand panel in \autoref{fig:default} shows that the mean instantaneous quality metric $T_{ij}$ of the active set ${\cal A}_i$ is systematically larger for merit-based sortition than for classical, random sortition. The difference is of the order of the standard deviation of the qualities of individual participants $\sigma_Q$, which is greater than the standard error on the mean by a factor of $\sqrt{N_{\rm act}}$. The high statistical significance of this difference demonstrates that the algorithm described in \S\ref{sec:method} fulfils the main goal of merit-based sortition.

\subsection{Large participant pool} \label{sec:large}
\begin{figure*}[tb]
\centering
\includegraphics[width=0.35\textwidth]{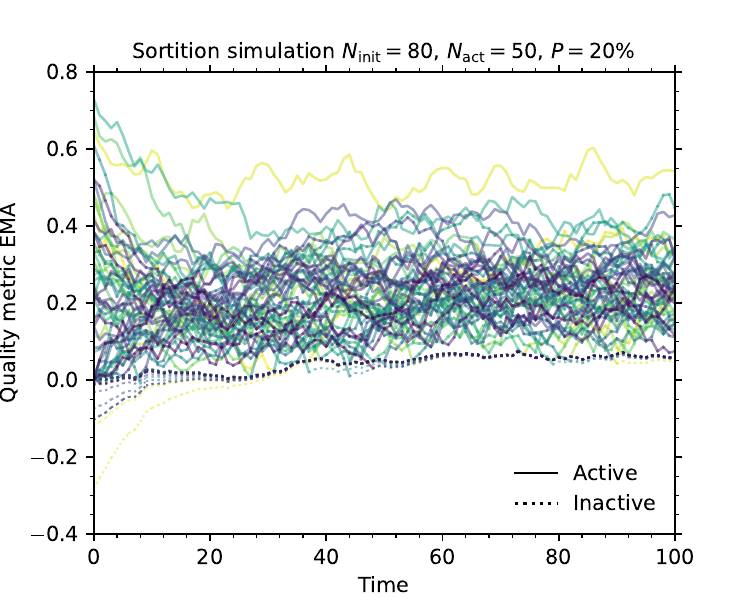}%
\hspace{-0.03\textwidth}
\includegraphics[width=0.35\textwidth]{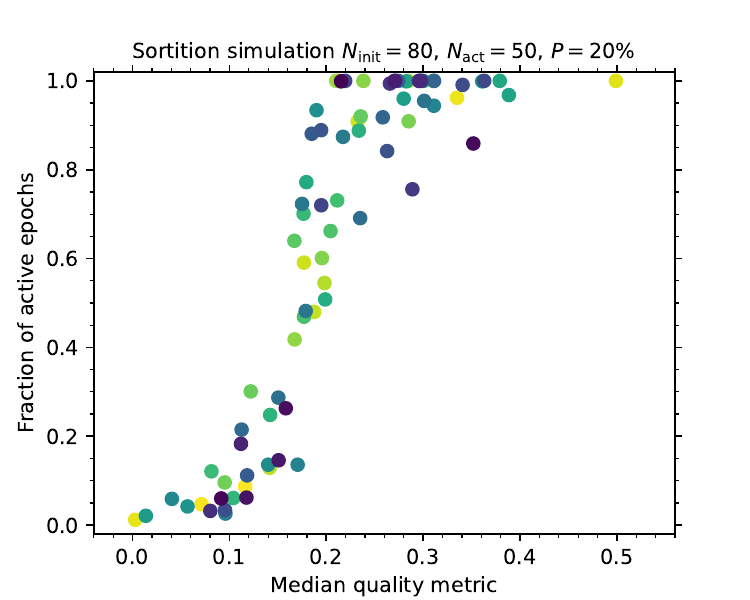}%
\hspace{-0.03\textwidth}
\includegraphics[width=0.35\textwidth]{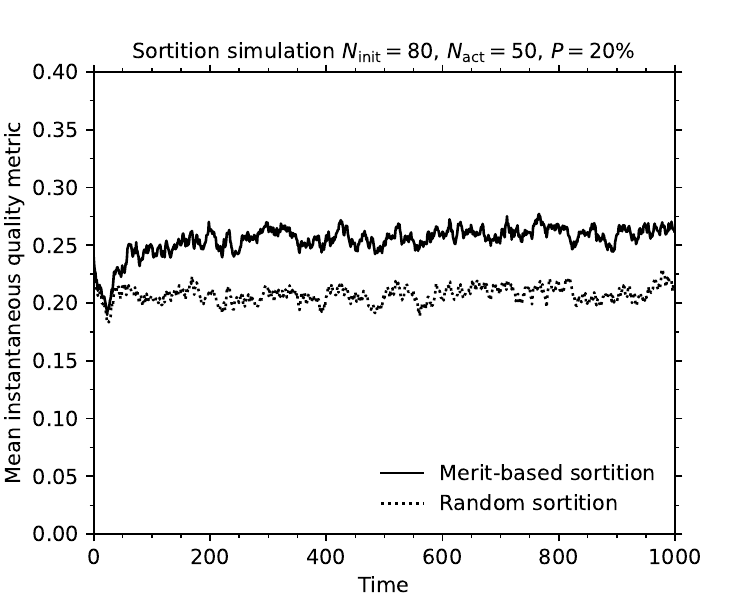}%
\caption{Merit-based sortition simulation for an initial set of 80 participants, illustrating the construction and performance of the active set ${\cal A}_i$ as a function of time in a crowded ecosystem. Left: Each color represents a different participant, and the line style reflects the (in)activity as indicated by the legend. Line segments connecting two epochs where a participant changes from inactive to active show as solid, whereas those connecting a transition from active to inactive show as dotted. Middle: Correlation between the median quality metric of each participant (reflecting their ability or performance) and the fraction of the 1000-epoch duration of the simulation for which this participant is active. Colors match those in the left-hand panel. Right: EMA of the mean instantaneous quality metric across the active set ${\cal A}_i$ for merit-based sortition (solid line) and random sortition (dotted line) over the full duration of the simulation. The panel titles list the number of initial participants, the number of active participants, and the percentile.\label{fig:large}}
\end{figure*}
\autoref{fig:large} illustrates how merit-based sortition handles participation in a more crowded ecosystem of $N_{\rm init}=80$ initial participants and an active set of $N_{\rm act}=50$ participants. The same mechanisms are at play as in \autoref{fig:default} -- inactive participants queue up at quality metric EMAs below the percentile reference value and get promoted successively whenever an active participant is relegated. Some examples of this promotion-relegation process are clearly visible at e.g.\ $i=\{37, 44, 79\}$. Due to the larger number of participants, the percentile is better defined and more stable, resulting in the formation of a crisp promotion-relegation threshold, visible as a set of clustered dotted lines in \autoref{fig:large}. This results in comparatively small differences in quality metric EMAs, but the ordering of participants is preserved. As a result, the rate of promotion and relegation is driven by the stochasticity of the instantaneous quality metrics $T_{ij}$ of the participants in the active set ${\cal A}_i$. If the participants' performance is stable compared to their mutual performance differences (i.e.\ $V<\sigma_Q$), then promotion and relegation proceed at a low rate, because only a subset of the participants has instantaneous quality metrics that scatter below the promotion-relegation threshold. However, if the participants' performance is comparatively volatile (i.e.\ $V>\sigma_Q$), then any participant in the active set might plausibly scatter below the promotion-relegation threshold, resulting in a larger rate of replacement. In the experiment shown here (with $V=0.2$ and $\sigma_Q=0.1$) we are looking at the latter case, with a relatively high rate of replacement. Higher $\alpha$ would increases the rate of replacement further.

As illustrated by comparing the middle panels of \autoref{fig:default} and \autoref{fig:large}, the positive correlation between participant performance and activity is sampled better with a larger participant pool, although the qualitative shape remains unchanged. The transition between mostly active and mostly inactive participants still takes place where the median quality metric is similar to $\mu_Q=0.2$. The greater degree of stability and predictability of the system manifests itself further in the right-hand panel of \autoref{fig:large}, which illustrates that the difference in the mean instantaneous quality metrics between experiments run with merit-based sortition and classical, random sortition are even more statistically significant than in \autoref{fig:default}. The factor-of-10 increase of the number of participants results in approximately a factor-of-3 decrease in the standard deviation of the mean, increasing the separability of both experiments and showing unambiguously that merit-based sortition results in a better-performing pool of active participants.

\subsection{Evolving participant pool} \label{sec:evolution}
\begin{figure*}[tb]
\centering
\includegraphics[width=0.35\textwidth]{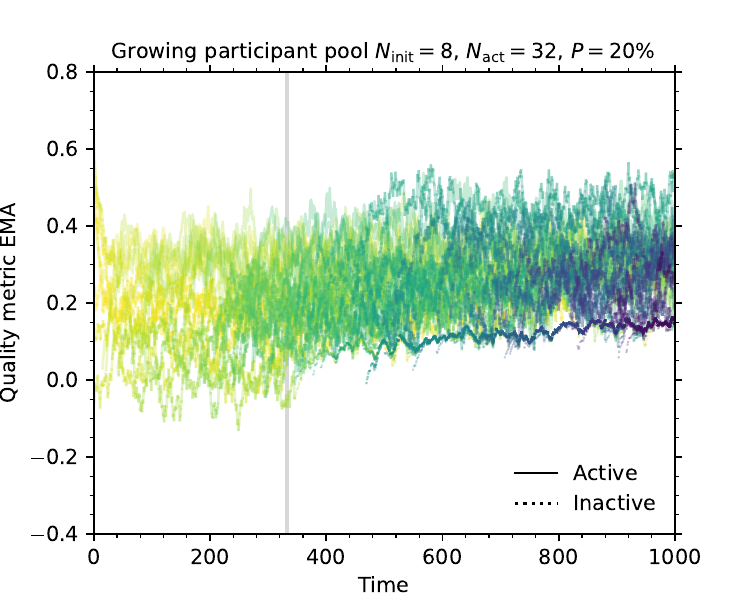}%
\hspace{-0.03\textwidth}
\includegraphics[width=0.35\textwidth]{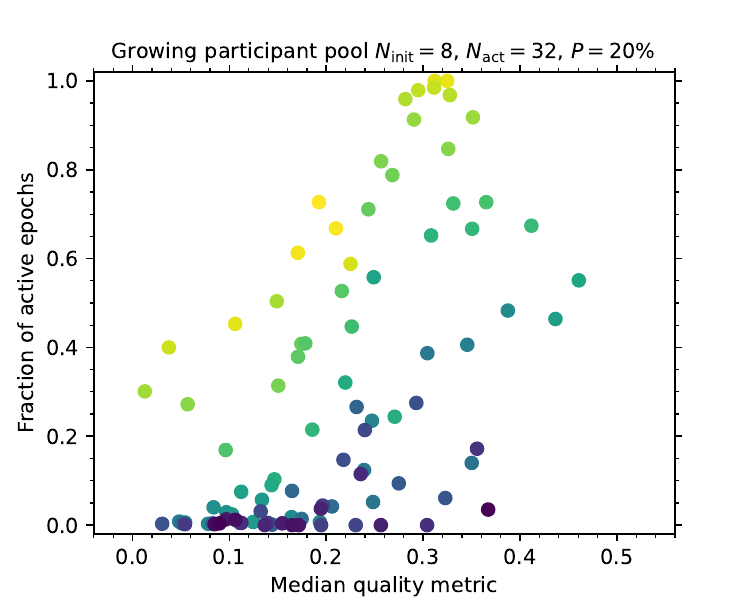}%
\hspace{-0.03\textwidth}
\includegraphics[width=0.35\textwidth]{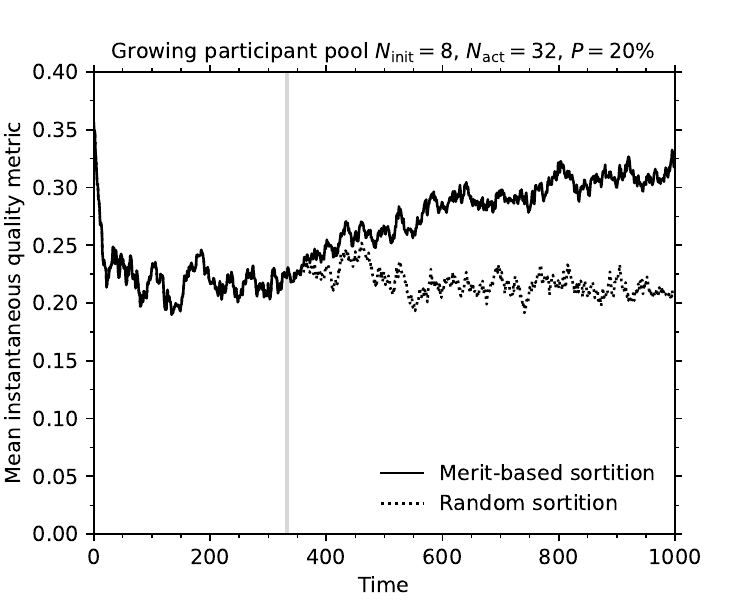}%

\includegraphics[width=0.35\textwidth]{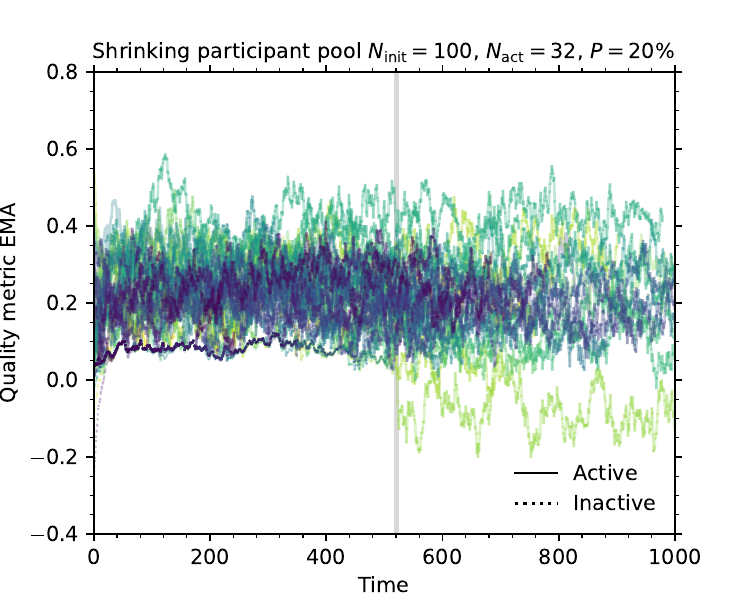}%
\hspace{-0.03\textwidth}
\includegraphics[width=0.35\textwidth]{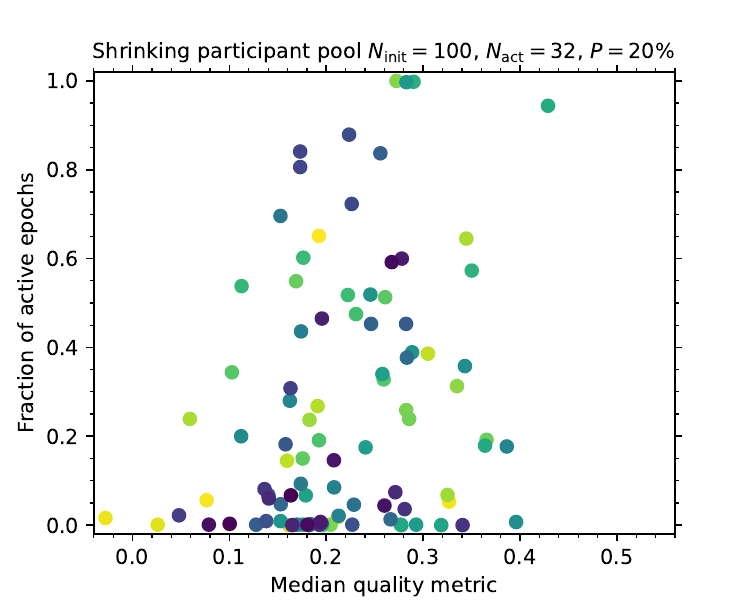}%
\hspace{-0.03\textwidth}
\includegraphics[width=0.35\textwidth]{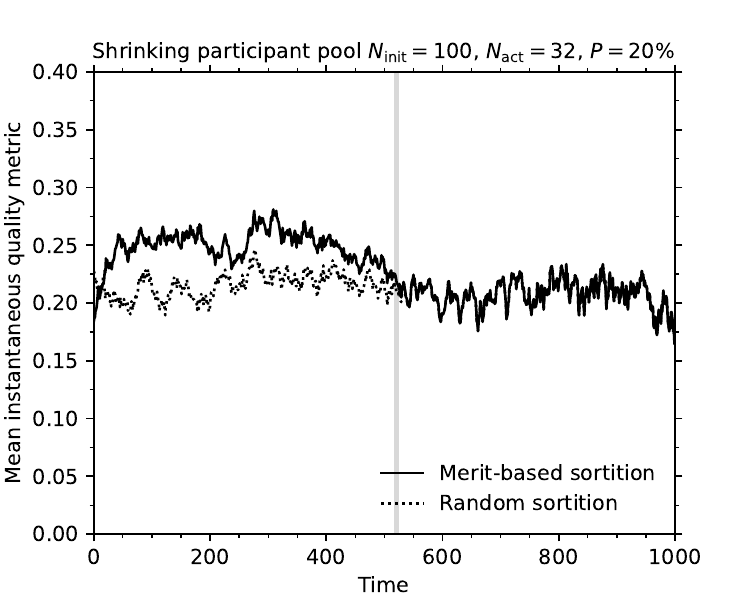}%

\includegraphics[width=0.35\textwidth]{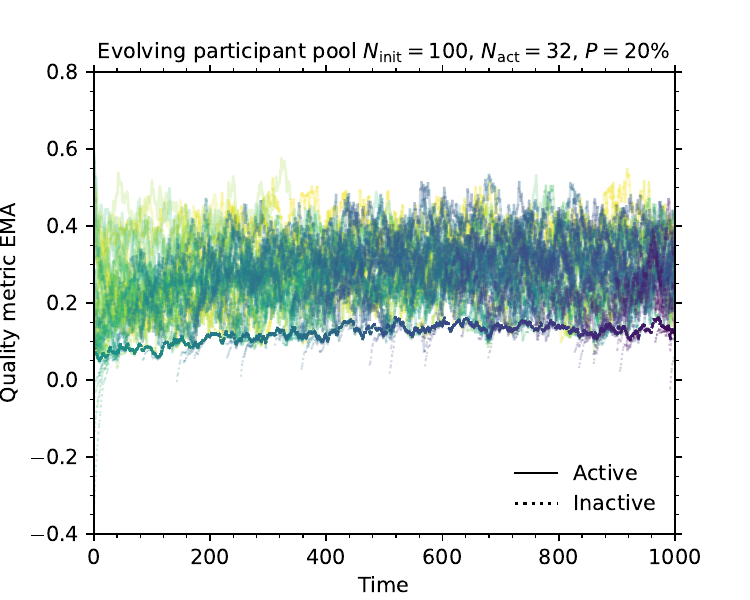}%
\hspace{-0.03\textwidth}
\includegraphics[width=0.35\textwidth]{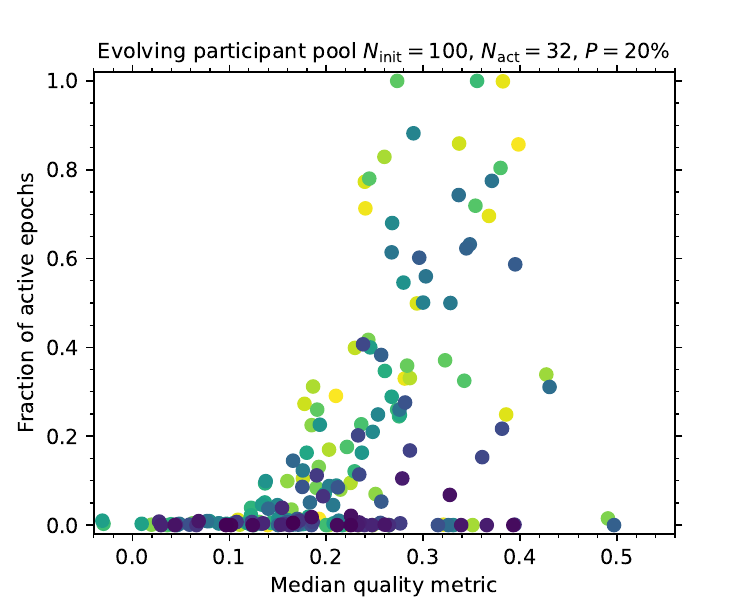}%
\hspace{-0.03\textwidth}
\includegraphics[width=0.35\textwidth]{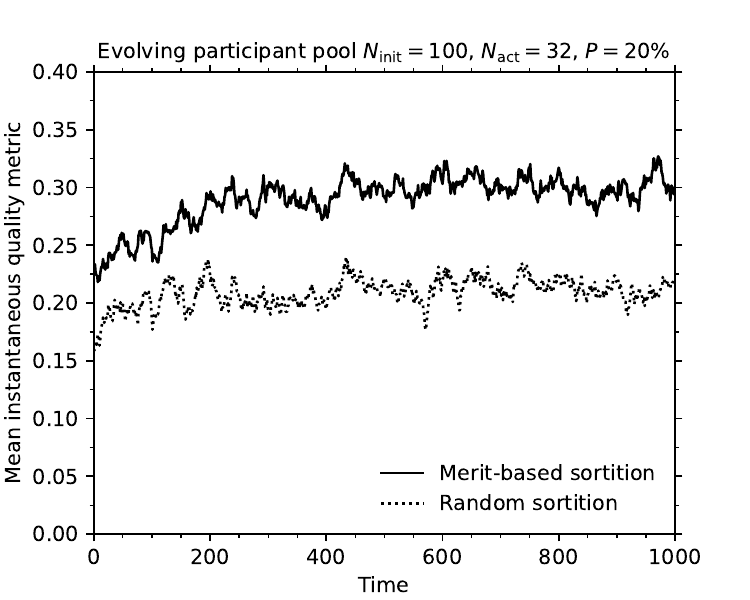}%
\caption{Merit-based sortition simulations for evolving participant pools. Colors indicate participants in order of their first participation (light to dark). Columns correspond to the same panels as in \autoref{fig:default} and \autoref{fig:large}.  The panel titles list the type of evolution (from top to bottom, these are growth, shrinkage, and dynamically evolving), the number of initial participants, the number of active participants, and the percentile.\label{fig:evolution}}
\end{figure*}
\autoref{fig:evolution} now expands on the idealized setup with a fixed number of participants, and instead considers different types of evolving total participant pools. Specifically, we consider three different setups:
\begin{enumerate}
\item
\textit{Growing participant pool}: The initial number of participants is $N_{\rm init}=8$ and new participants join the inactive set ${\cal I}_i$ at a rate defined by $p_{\rm growth}=10^{-1}$. They are initialized according to \autoref{eq:q_target_inactive}. None of the participants ever leave. After 1000 epochs, the participant pool has grown to nearly 100 participants.
\item
\textit{Shrinking participant pool}: The initial number of participants is $N_{\rm init}=100$ and existing participants leave the participant pool, independently of whether or not they are active, at a rate defined by $p_{\rm attr}=2\times10^{-3}$. No new participants ever join. After 1000 epochs, the participant pool has shrunk to around 10 participants.
\item
\textit{Evolving participant pool}: The initial number of participants is $N_{\rm init}=100$ and participants constantly join and leave, at rates identical to the previous two cases, i.e.\ $p_{\rm growth}=10^{-1}$ and $p_{\rm attr}=2\times10^{-3}$. This results in a variable number of participants, ranging between 90 and 110.
\end{enumerate}

As shown by \autoref{fig:evolution}, the added value of merit-based sortition appears only when $N_{\rm tot}>N_{\rm act}$. When the participant pool size grows beyond or shrinks below $N_{\rm act}$, there is an immediate change in the dispersion of the quality metric EMA, as merit-based sortition cuts the lowest instantaneous quality metrics. As indicated in \S\ref{sec:large}, the larger the participant pool is, the clearer the promotion-relegation threshold becomes.

Due to the evolving nature of the participant pool, the relatively well-defined correlation between participant performance and activity (middle column) is erased once we allow growth and attrition of participants. In dynamically evolving participant pools, the fraction of the total number of epochs that a participant is active is strongly influenced by the fraction of time they participate. However, as indicated by the trends visible among similar colors in the top-middle panel of \autoref{fig:evolution} (growth without attrition), which represent the order in which the participant appeared, the correlation persists among participants with similar moments of appearance. This indicates that the fraction of a participant's \textit{participation} time interval during which it is active does correlate with its performance. Any correlation between the quality metric and total activity fraction vanishes altogether for non-zero attrition, as the broad distribution of participation time intervals outweighs the fraction of that time interval that the participant is active. The short participation time interval is most clearly illustrated in the bottom-middle panel, where a large number of participants is active for a near-zero fraction of the total number of epochs. As stated, this is driven by their relatively short-lived participation.

The right-hand column in \autoref{fig:evolution} illustrates that the added value of merit-based sortition increases towards larger $N_{\rm tot}/N_{\rm act}$. This manifests itself through a larger difference between the solid and dotted lines for more populous systems (see the left-hand column). It can be understood as a size-of-sample effect: for an increasing number of participants, the active set is restricted to an ever-smaller tail end of the distribution, which implies that the mean instantaneous quality metric is higher than the average across the entire pool. The initial growth of the difference between the solid and dotted lines arises from the initialization of the EMA, and subsides after a brief equilibriation phase. Finally, and perhaps most importantly, we see that the difference between merit-based sortition and classical, random sortition is greater for dynamically evolving participant pools than for a fixed participant pool (compare \autoref{fig:large} and \autoref{fig:evolution}). This affirms the need for merit-based sortition in decentralized systems.

\subsection{Choice of percentile} \label{sec:percentiles}
\begin{figure*}[tb]
\centering
\includegraphics[width=0.4\textwidth]{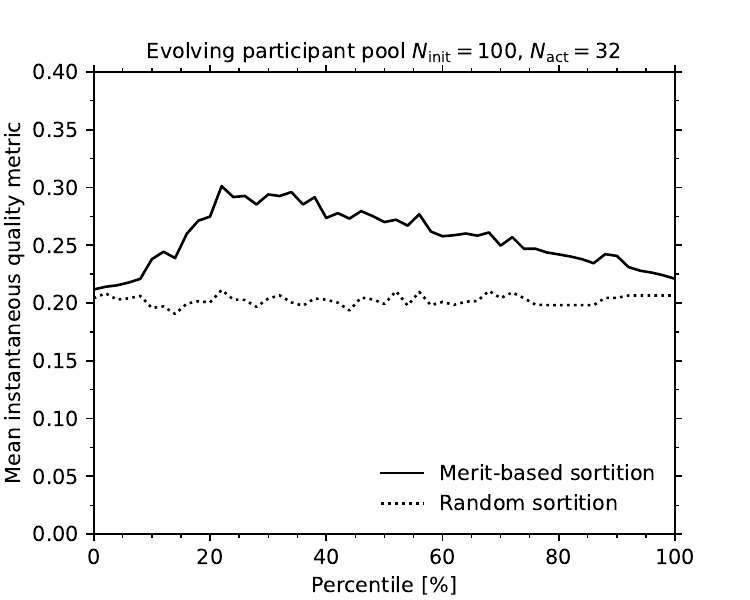}%
\includegraphics[width=0.4\textwidth]{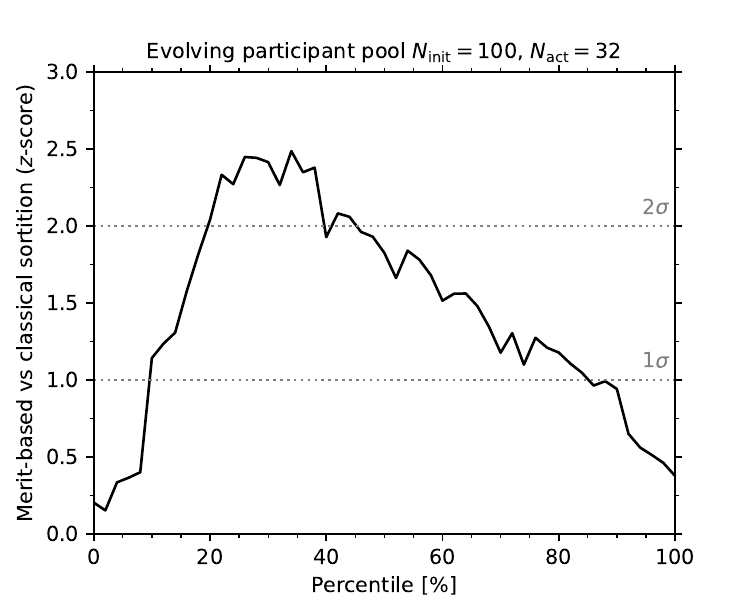}%
\caption{Set of sortition simulations showing how the added value of merit-based sortition over classical, random sortition depends on the percentile $P$, for a dynamically evolving participant pool. Left: Time-averaged mean instantaneous quality metric ${\cal T}_i$ across the active set ${\cal A}_i$ as a function of $P$. The solid line indicates the results for merit-based sortition, and the dotted line represents classical (random) sortition. Right: Statistical $z$-score (the absolute difference between both lines in the left panel, normalized by the standard deviation of the mean underlying, unsmoothed instantaneous quality metrics) as a function of $P$, demonstrating a highly statistically significant ($>2\sigma$) improvement with merit-based sortition relative to classical sortition for $P=20{-}40\%$, and a mild ($>1\sigma$) improvement for $P=10{-}85\%$.
\label{fig:percentiles}}
\end{figure*}
They key question now is how the percentile $P$ should be chosen -- this is the main free parameter of the merit-based sortition process. We repeat the experiment of the bottom row of \autoref{fig:evolution} for 50 different values of the percentile evenly spread over the interval $P=0{-}100\%$. The resulting mean instantaneous quality metrics across the active set (shown in the bottom-right panel of \autoref{fig:evolution} as a function of time) are reduced to a single value per experiment by additionally taking the mean across all epochs. The resulting relation between the mean instantaneous quality metric (taking the mean over ${\cal A}_i$ and over time) is shown as a function of $P$ in \autoref{fig:percentiles} for merit-based sortition and classical, random sortition.

\autoref{fig:percentiles} shows that the added value of merit-based sortition is smallest at the extremes of the percentile $P$. This is unsurprising -- at small percentiles, the depth and rate of replacement are small, such that most participants who are initially part of the active set remain. As a result, merit-based sortition can only provide a minor boost to the mean instantaneous quality metric of the active set. A similar outcome is found at high percentiles, where nearly all participants of the active set are continuously being replaced. As a result, the active set is populated almost randomly and the difference relative to random sortition is small, but non-zero thanks to the ordering by performance caused by the quality metric EMA.

In summary, merit-based sortition adds value when the fraction of participants that is replaced is sufficiently large to enable selecting the best ones, but also is sufficiently small so that high-quality participants remain in the active set. These requirements directly imply that there exists an optimal percentile at which the instantaneous quality metric of participants in the active set is maximized. \autoref{fig:percentiles} shows that this occurs at a statistical significance of $>2\sigma$ in the range $P=20{-}40\%$ for the experiments performed here (see the right-hand panel of \autoref{fig:percentiles} -- a $>1\sigma$ improvement is achieved for $P=10{-}85\%$). The precise optimum may vary based on the quantitative distribution function of the participants' median quality metrics $\tilde{Q}_j$. We find that $P=25\%$ provides a good starting point for fine-tuning experiments.

\section{Discussion and Conclusion} \label{sec:disc}
We propose a simple mechanism for merit-based sortition, which employs an exponential moving average (EMA) of a quality metric to robustly predict the expected performance during the current epoch. The EMA of inactive participants is updated using a chosen percentile of instantaneous (unsmoothed) quality metrics among the active set. This mechanism allows the selection of active participants from a much larger sample based on the expected quality of their work. This merit-based sortition scheme has a variety of favorable properties:
\begin{enumerate}
    \item The active set of participants ${\cal A}_i$ is permeable and enables participants from the inactive set ${\cal I}_i$ to enter and prove their worth. This permeability allows the algorithm to appropriately handle quality changes among the participants.
    \item The fraction of the active set that is replaced and the frequency with which this occurs is controlled by a single parameter $0<P[\%]\leq100$, which represents the percentile of the quality metric among the active set that is used to update the metric's EMA of inactive participants. High $P$ implies a high depth and rate of replacement.
    \item Participants who consistently perform above the $(100\%-P)$-th percentile of the active set are not at risk of being replaced and remain active. The consistency is critical, as the volatility of their performance must be sufficiently small to avoid incidental, poor performances that risk relegation.
    \item Participants performing below the $P$-th percentile of the active set are at risk of being replaced.
    \item The probability of an inactive participant to be promoted into the active set increases with its prior EMA-smoothed quality metric $Q_{i-1,j}$, the volatility $V$ of instantaneous quality metrics in the active set $T_{ij}$, the smoothing factor $\alpha$, and the percentile $P$.
    \item The time required by relegated participants to return to the active set increases for more poorly performing participants, because these are separated further from the minimum quality metric among the active participants.
    \item It is possible to define an optimal percentile at which the mean instantaneous quality metric across the active set ${\cal T}_i$ is maximized. For the experiments performed here, this is found to fall in the range $P=20{-}40\%$ at $>2\sigma$ statistical significance, but the precise optimum may vary based on the quantitative distribution function of the participants' median quality metrics. $P=25\%$ provides a good initial value.
\end{enumerate}

The above properties of the merit-based sortition scheme presented in this work are favorable, because they demonstrate that participants are involved in an active set based on the quality of their work. It shares the property of classical, random sortition that it draws from a much larger pool of participants, of which each individual has a non-zero probability to become part of the active set. As a result, for an infinite number of epochs each participant is guaranteed to have been part of the active set at least once. However, the key difference is that the probability of inclusion in the active set is variable and merit-based. As such, the mechanism outlined here guarantees equality of opportunity instead of equality of outcome, and thereby fulfils a key requirement for well-functioning decentralized systems.

\begin{sloppypar}
\bibliographystyle{wp}
{\small
\bibliography{ourbib}

\begin{thebibliography}{17}
\expandafter\ifx\csname natexlab\endcsname\relax\def\natexlab#1{#1}\fi

\bibitem[{Banerjee {et~al.}(2023)Banerjee, Ramanathan, \& Kumar}]{banerjee23}
Banerjee, S., Ramanathan, N., \& Kumar, R. 2023, in 2023 IEEE International
  Conference on Omni-layer Intelligent Systems (COINS), 1--7.
\href{https://doi.org/10.1109/COINS57856.2023.10189237}{https://doi.org/10.1109/COINS57856.2023.10189237}\vspace{-3pt}

\bibitem[{Breidenbach {et~al.}(2021)Breidenbach, Cachin, Chan, Coventry, Ellis,
  Juels, Koushanfar, Miller, Magauran, Moroz, Nazarov, Topliceanu, Tram\`er, \&
  Zhang}]{breidenbach21}
Breidenbach, L., Cachin, C., Chan, B., {et~al.} 2021, Chainlink 2.0: Next Steps
  in the Evolution of Decentralized Oracle Networks.
\href{https://research.chain.link/whitepaper-v2.pdf}{https://research.chain.link/whitepaper-v2.pdf}\vspace{-3pt}

\bibitem[{Buterin(2014)}]{buterin14}
Buterin, V. 2014, Ethereum: A next-generation smart contract and decentralized
  application platform.
\href{https://ethereum.org/content/whitepaper/whitepaper-pdf/Ethereum\_Whitepaper\_-\_Buterin\_2014.pdf}{https://ethereum.org/content/whitepaper/whitepaper-pdf/Ethereum\_Whitepaper\_-\_Buterin\_2014.pdf}\vspace{-3pt}

\bibitem[{Ellis {et~al.}(2017)Ellis, Juels, \& Nazarov}]{ellis17}
Ellis, S., Juels, A., \& Nazarov, S. 2017, ChainLink: A Decentralized Oracle
  Network.
\href{https://research.chain.link/whitepaper-v1.pdf}{https://research.chain.link/whitepaper-v1.pdf}\vspace{-3pt}

\bibitem[{Engelstad(1989)}]{engelstad89}
Engelstad, F. 1989, Social Science Information, 28, 23.
\href{https://doi.org/10.1177/053901889028001002}{https://doi.org/10.1177/053901889028001002}\vspace{-3pt}

\bibitem[{Flanigan {et~al.}(2021)Flanigan, G{\"o}lz, Gupta, Hennig, \&
  Procaccia}]{flanigan21}
Flanigan, B., G{\"o}lz, P., Gupta, A., Hennig, B., \& Procaccia, A.~D. 2021,
  Nature, 596, 548.
\href{https://doi.org/10.1038/s41586-021-03788-6}{https://doi.org/10.1038/s41586-021-03788-6}\vspace{-3pt}

\bibitem[{Freitas {et~al.}(2023)Freitas, Tonkikh, Bendoukha,
  Tucci-Piergiovanni, Sirdey, Stan, \& Kuznetsov}]{freitas23}
Freitas, L., Tonkikh, A., Bendoukha, A.-A., {et~al.} 2023, Homomorphic
  Sortition -- Secret Leader Election for PoS Blockchains, arXiv:2206.11519.
\href{https://arxiv.org/abs/2206.11519}{https://arxiv.org/abs/2206.11519}\vspace{-3pt}

\bibitem[{Gilad {et~al.}(2017)Gilad, Hemo, Micali, Vlachos, \&
  Zeldovich}]{gilad17}
Gilad, Y., Hemo, R., Micali, S., Vlachos, G., \& Zeldovich, N. 2017, in
  Proceedings of the 26th Symposium on Operating Systems Principles, SOSP '17
  (New York, NY, USA: Association for Computing Machinery), 51--68.
\href{https://doi.org/10.1145/3132747.3132757}{https://doi.org/10.1145/3132747.3132757}\vspace{-3pt}

\bibitem[{{Headlam-Morley}(1891)}]{headlam1891}
{Headlam-Morley}, J.~W. 1891, Election by Lot at Athens, 1st edn. (Cambridge:
  Cambridge University Press).
\vspace{-3pt}

\bibitem[{Jacquet {et~al.}(2022)Jacquet, Niessen, \& Reuchamps}]{jacquet22}
Jacquet, V., Niessen, C., \& Reuchamps, M. 2022, International Political
  Science Review, 43, 295.
\href{https://doi.org/10.1177/0192512120949958}{https://doi.org/10.1177/0192512120949958}\vspace{-3pt}

\bibitem[{Kruijssen {et~al.}(2024)Kruijssen, Emmons, Peluso, Ghaffar, Huang,
  Longmore, \& Kell}]{kruijssen24}
Kruijssen, J. M.~D., Emmons, N., Peluso, K., {et~al.} 2024, Allora
  Decentralized Intelligence, 1, 1.
\href{https://doi.org/10.70235/allora.0x10001}{https://doi.org/10.70235/allora.0x10001}\vspace{-3pt}

\bibitem[{Nakamoto(2008)}]{nakamoto08}
Nakamoto, S. 2008, Bitcoin: A Peer-to-Peer Electronic Cash System,
  https://bitcoin.org/bitcoin.pdf.
\href{https://doi.org/10.2139/ssrn.3440802}{https://doi.org/10.2139/ssrn.3440802}\vspace{-3pt}

\bibitem[{Rao {et~al.}(2021)Rao, Steeves, Shaabana, Attevelt, \&
  McAteer}]{rao21}
Rao, Y., Steeves, J., Shaabana, A., Attevelt, D., \& McAteer, M. 2021,
  BitTensor: A Peer-to-Peer Intelligence Market.
\href{https://doi.org/10.48550/arXiv.2003.03917}{https://doi.org/10.48550/arXiv.2003.03917}\vspace{-3pt}

\bibitem[{Saa \& Stern(2019)}]{saa19}
Saa, O. \& Stern, J.~M. 2019, Proceedings, 33, 17.
\href{https://doi.org/10.3390/proceedings2019033017}{https://doi.org/10.3390/proceedings2019033017}\vspace{-3pt}

\bibitem[{Sintomer(2023{\natexlab{a}})}]{sintomer23}
Sintomer, Y. 2023{\natexlab{a}}, The Disappearance of Sortition in Politics: A
  Historical Enigma (Cambridge University Press), 125--187.
\href{https://doi.org/10.1017/9781009285650.004}{https://doi.org/10.1017/9781009285650.004}\vspace{-3pt}

\bibitem[{Sintomer(2023{\natexlab{b}})}]{sintomer23b}
Sintomer, Y. 2023{\natexlab{b}}, Sortition and Politics in the Twenty-First
  Century (Cambridge University Press), 250--278.
\href{https://doi.org/10.1017/9781009285650.006}{https://doi.org/10.1017/9781009285650.006}\vspace{-3pt}

\bibitem[{W{\"u}st \& Gervais(2018)}]{wuest18}
W{\"u}st, K. \& Gervais, A. 2018, in 2018 Crypto Valley Conference on
  Blockchain Technology (CVCBT), 45--54.
\href{https://doi.org/10.1109/CVCBT.2018.00011}{https://doi.org/10.1109/CVCBT.2018.00011}\vspace{-3pt}

\end{thebibliography}
}
\end{sloppypar}

\end{document}